\documentclass[aps,prb,twocolumn,groupedaddress,showpacs,floatfix]{revtex4}
\usepackage{graphicx}
\def\be{\begin{equation}}
\def\ee{\end{equation}}
\def\ba{\begin{eqnarray}}
\def\ea{\end{eqnarray}}
\def\bc{\begin{center}}
\def\ec{\end{center}}

\begin{document}

\title{Microwave response of a two-dimensional electron stripe}

\author{S. A. Mikhailov}
\email[E-mail address: ]{sergey.mikhailov@mh.se}
\author{N. A. Savostianova}
\affiliation{Mid-Sweden University, ITM, Electronics Design Division, 851 70 Sundsvall, Sweden}

\date{\today}

\begin{abstract}
Electromagnetic response of a finite-width two-dimensional electron stripe is theoretically studied. It is shown that retardation and radiative effects substantially modify the absorption spectrum of the system at microwave frequencies, leading to a non-trivial zigzag behavior of the magnetoplasmon-polariton modes in magnetic fields, similar to that recently observed by Kukushkin et al.  [Phys. Rev. Lett. {\bf 90}, 156801 (2003)].
\end{abstract}

\pacs{73.20.Mf, 73.21.-b, 71.36.+c, 78.70.Gq}








\maketitle

\section{\label{intro}Introduction}

Plasma oscillations in low-dimensional electron systems (ES), -- inversion layers, systems on the surface of liquid helium, quantum wells, wires and dots, -- were experimentally studied since 1976 \cite{Grimes76,Allen77,Theis78} and seemed to be well understood (see e.g. \cite{Theis80,Allen83,Heitmann86,Demel88,Demel90,Demel91,Heitmann92a,Heitmann93}). Far-infrared transmission, as well as Raman spectroscopy experiments quantitatively confirmed theoretically predicted \cite{Stern67,Chiu74} plasmon and magnetoplasmon spectra, both in two-dimensional (2D) electron layers, and in systems with lower dimensionality (wires and dots). However, recent experimental studies \cite{Kukushkin03a,Kukushkin03b} of the microwave response of macroscopic 2D electron disks with the diameter $\simeq 1$ mm revealed very interesting and fully unexpected features of the magnetoplasmon spectra. Response of relatively small disks (with the diameter $\lesssim 0.1$ mm) was very similar to that of microscopic  dots. The spectrum consisted of two sharp resonances corresponding to the bulk and edge magnetoplasmon modes with the frequencies
$\omega_\pm(B)=\sqrt{\omega_p^2+\omega_c^2/4}\pm\omega_c/2$,
where $\omega_c$ is the cyclotron frequency and $\omega_p$ the plasma frequency at zero magnetic field $B=0$. In larger samples, however, the upper (bulk-magnetoplasmon) mode exhibited substantial changes. At a certain finite value of $B$ this mode intersected the cyclotron resonance (CR) line $\omega=\omega_c$ and disappeared experiencing a dramatic broadening. A second, and in some cases a third mode, with comparable or even larger amplitude, appeared in the absorption spectrum at higher frequencies and showed similar behavior. As a result, the $B$-field dependence of the bulk magnetoplasma modes exhibited intriguing zigzag behavior as a function of magnetic field at $\omega>\omega_p$ \cite{Kukushkin03a,Kukushkin03b}.

It was shown in \cite{Kukushkin03a,Kukushkin03b} that the observed features of the magnetoplasmon spectra are due to the influence of retardation effects. At $B=0$ the measured resonance frequencies turned out to be in a good quantitative agreement with Ref.  \cite{Stern67}, and at finite $B$ calculations did reveal a certain zigzag behavior of the resonant bulk magnetoplasmon modes. However the physical reasons of such a strange behavior of the modes at finite $B$, as well as the overall behavior of the linewidth of resonances, have not been explained. 

The goal of this paper is to theoretically study electromagnetic response of a finite-size macroscopic 2DES. We consider a 2D stripe of a finite width $W$ and focus on the regime, where the retardation and radiative effects become important \cite{Mikhailov04a} (for mm-size 2D samples this corresponds to microwave frequencies -- the parameter range where very interesting microwave experiments \cite{Mani02,Zudov03,Dorozhkin03,Yang03,Mani04,Studenikin04,Kukushkin04a,Willett03} have been recently done). Although the stripe geometry differs from the experimental one (disks), our calculations reproduce all the experimental findings \cite{Kukushkin03a,Kukushkin03b} and allow one to understand the physics of the observed effects. It will be shown that the unusual zigzag behavior of the bulk-magnetoplasmon modes (more exactly, magnetoplasmon-polaritons) is explained by an interplay of radiative and dissipative contributions to their linewidth. 

In an infinite sample the spectrum of 2D magnetoplasmons has been considered (taking into account retardation but ignoring the damping of modes) by Chiu and Quinn \cite{Chiu74} in 1974. Their results are applicable for the description of far-infrared response, when the size of the sample is large as compared to the magnetoplasmon and radiation wavelength, and the issue of linewidth is not very important. We consider a finite-size geometry, more relevant in view of the recent microwave experiments \cite{Kukushkin03a,Kukushkin03b,Mani02,Zudov03,Dorozhkin03,Yang03,Mani04,Studenikin04,Kukushkin04a,Willett03}, where the radiation and the plasmon wavelengths exceed or comparable with the sample dimensions, and where the linewidth of modes essentially determines the observed non-trivial features of the absorption spectra.

\section{\label{form}theory}

Assume that the 2D electron-gas stripe ($-W/2<x<W/2$, $-\infty<y<\infty$) lies in the plane $z=0$, the external magnetic field ${\bf B}=(0,0,B)$ is directed along the $z$-axis, and the dielectric constant $\epsilon$ of surrounding medium is uniform in all the space. Let an external plane electromagnetic wave with the frequency $\omega$ and the electric field ${\bf E}^{ext}({\bf r},t)={\bf E}^{ext}e^{i\omega(z\sqrt{\epsilon}/c- t)}$ be normally incident upon the structure. It induces in the 2D system oscillating charge, current ${\bf j}$, and -- around the 2DES -- electric and magnetic fields. The total electric field ${\bf E}^{tot}={\bf E}^{ext}+{\bf E}^{ind}$, given by the sum of the external and the scattered (induced) fields, satisfies the Maxwell equation
\begin{equation}
\nabla \times (\nabla \times {\bf E}^{\rm tot}) + \frac {\epsilon} {c^2}
\frac {\partial^2 {\bf E}^{\rm tot}} {\partial t^2} = - \frac {4\pi}{c^2}
\frac{\partial}{\partial t}{\bf j}(x,t)\delta(z).
\label{me}
\end{equation} 
Looking for a solution, proportional to $e^{-i\omega t}$, and assuming that fields and currents do not depend on the $y$-coordinate, we rewrite Eq. (\ref{me}) in an integral form for the $x$- and $y$-components of the induced electric field
\be
E_x^{ind}(x)=-\frac i{\omega\epsilon}\int_ {-\infty}^{\infty}dq_x\kappa e^{iq_xx}\int_{-W/2}^{W/2}e^{-iq_xx'}j_x(x')dx',
\label{ex}
\ee
\be
E_y^{ind}(x)=\frac {i\omega}{c^2}\int_ {-\infty}^{\infty}\frac{dq_x}\kappa e^{iq_xx}\int_{-W/2}^{W/2}e^{-iq_xx'}j_y(x')dx'.
\label{ey}
\ee
Here $\kappa=\sqrt{q_x^2-\omega^2\epsilon/c^2}$ determines the $z$-dependence of fields ($\propto e^{-\kappa|z|}$). Dependent on the interval of the $dq_x$-integration, Eqs. (\ref{ex})--(\ref{ey}) describe both the evanescent induced fields (at $q_x^2>\omega^2\epsilon/c^2$) and the scattered outgoing wave (at $q_x^2<\omega^2\epsilon/c^2$); the sign of Im $\kappa$ should be properly chosen to satisfy the scattering boundary conditions at $|z|\to\infty$.

Now we need to relate the current and the total electric field inside the stripe. We do not consider in this paper possible nonlocal effects and use the relation $j_\alpha ({\bf r})=\sigma_{\alpha\beta}E^{tot}_\beta ({\bf r})\theta(W/2-|x|) $, where $\sigma_{\alpha\beta}$ is the conductivity tensor of the 2DES, dependent on the frequency, magnetic field and the momentum relaxation rate, and $\theta(x)$ is the step function. Together with Eqs. (\ref{ex})--(\ref{ey}) this gives a system of two integral equations for the electric current inside the stripe $|x|<W/2$,
\begin{widetext}
\be
j_x(x)=\sigma_{x\alpha}E_\alpha^{ext}-\frac {i\sigma_{xx}}{\omega\epsilon}\int_ {-\infty}^{\infty}dq_x\kappa e^{iq_xx}\int_{-W/2}^{W/2}e^{-iq_xx'}j_x(x')dx'
+
\frac {i\omega\sigma_{xy}}{c^2}\int_ {-\infty}^{\infty}\frac{dq_x}\kappa e^{iq_xx}\int_{-W/2}^{W/2}e^{-iq_xx'}j_y(x')dx',
\ee
\be
j_y(x)=\sigma_{y\alpha}E_\alpha^{ext}-\frac {i\sigma_{yx}}{\omega\epsilon}\int_ {-\infty}^{\infty}dq_x\kappa e^{iq_xx}\int_{-W/2}^{W/2}e^{-iq_xx'}j_x(x')dx'
+
\frac {i\omega\sigma_{yy}}{c^2}\int_ {-\infty}^{\infty}\frac{dq_x}\kappa e^{iq_xx}\int_{-W/2}^{W/2}e^{-iq_xx'}j_y(x')dx'.
\ee
\end{widetext}
These equations are solved by reduction to a matrix form. Expanding the currents in a complete set of functions,

\be
\left(
\begin{array}{c}
j_x(x)\\
j_y(x)
\end{array}\right)
=\sum_{n=0}^\infty 
\left(
\begin{array}{c}
A_n\\
B_n
\end{array}\right)
\cos\left(\frac{2\pi x}W\left(n+\frac 12\right)\right),\label{jxy}
\ee
and following the standard procedure we get the following system of algebraic equations
\be
\sum_{n=0}^\infty M^{(11)}_{mn}A_n +\sum_{n=0}^\infty M^{(12)}_{mn}B_n = \frac{2(-1)^m}{\pi(m+1/2)}\sigma_{x\alpha}E_\alpha^{ext},
\label{a1}
\ee
\be
\sum_{n=0}^\infty M^{(21)}_{mn}A_n +\sum_{n=0}^\infty M^{(22)}_{mn}B_n = \frac{2(-1)^m}{\pi(m+1/2)}\sigma_{y\alpha}E_\alpha^{ext},
\label{a2}
\ee
where

\be
M^{(11)}_{mn}=\delta_{mn}+\frac {8\pi^2i\sigma_{xx}}{\omega\epsilon W}
J_{mn}^{(+)}(\Omega),
\label{m11}
\ee
\be
M^{(12)}_{mn}=-
\frac {2i\omega\sigma_{xy}W}{c^2}
J_{mn}^{(-)}(\Omega),
\label{m12}
\ee
\be
M^{(21)}_{mn}=\frac {8\pi^2i\sigma_{yx}}{\omega\epsilon W}
J_{mn}^{(+)}(\Omega),
\label{m21}
\ee
\be
M^{(22)}_{mn}=\delta_{mn}-
\frac {2i\omega\sigma_{yy}W}{c^2}
J_{mn}^{(-)}(\Omega),
\label{m22}
\ee
and $\Omega=\omega W\sqrt{\epsilon}/2\pi c$. The integrals $J_{mn}^{(\pm)}$ are defined as

\be
J_{mn}^{(\pm)}(\Omega)=
\int_ {-\infty}^{\infty} dQ (Q^2-\Omega^2)^{\pm 1/2} \gamma_m(Q) \gamma_n(Q),
\label{intg}
\ee
where 
\be
\gamma_n(Q)=\frac{(-1)^{n+1}(n+1/2)}{\pi}\frac{\cos\pi Q}{Q^2-(n+1/2)^2}.
\ee
The part of the integrals in (\ref{intg}) from $Q=-\Omega$ to $Q=\Omega$ gives imaginary contribution to the matrix elements in (\ref{a1})--(\ref{a2}) and physically corresponds to the radiative decay.

Having solved equations (\ref{a1})--(\ref{a2}) and having got the coefficients $A_n$, $B_n$ in terms of the external field $E_\alpha^{ext}$, we can calculate the current, Eq. (\ref{jxy}), and the fields, Eqs. (\ref{ex})--(\ref{ey}), inside the stripe. Then, calculating the Joule heat $Q=\int_{-W/2}^{W/2}dx ({\bf j}^*{\bf E}+{\bf j}{\bf E}^*)/4$ and dividing it by the energy flow $S_z=c\sqrt{\epsilon}(|E_x^{ext}|^2+|E_y^{ext}|^2)/8\pi$, we get the  absorption cross section $Q/S_z$ (dimensionality cm). All plots of the next Section show the frequency, magnetic field and other dependencies of the dimensionless ``absorption coefficient'' 
\be
A=\frac Q{S_zW},\label{A}
\ee
defined as the absorption cross section $Q/S_z$ normalized by the width of the sample.

\section{\label{res}results}
\subsection{Preliminary notes}

The plasma frequency in a 2DES, calculated in the quasi-static approximation is given by the formula \cite{Stern67}
\be
\omega_p^2(q)=\frac{2\pi n_se^2}{m^\star\epsilon}q,
\label{2Dp}
\ee
where $n_s$, $e$ and $m^\star$ are the density, the charge and the effective mass of 2D electrons, and $q$ is the plasmon wavevector. In a stripe of a finite width $W$ the lowest quasistatic eigen-mode frequency is expected at $q\approx \pi/W$. Therefore we choose for the frequency unit the quantity
\be
\omega_0=\sqrt{\frac{2\pi^2 n_se^2}{m^\star\epsilon W}};
\label{omega0}
\ee
for a typical macroscopic (mm-size) GaAs 2D sample its value lies in the tens-of-gigahertz range,
\be
\frac{\omega_0}{2\pi}=43.3\times \sqrt{\frac{n_s[10^{11}\ {\rm cm}^{-2}]}{\epsilon W[{\rm mm}]}}\ [{\rm GHz}];
\ee
(for the effective mass we use $m^\star/m_0=0.067$). Everywhere below we measure the frequencies in units $\omega/\omega_0$, magnetic fields in units $\omega_c/\omega_0$, and the momentum relaxation rate $\gamma=e/m^\star\mu$ in units $\gamma/\omega_0$ ($\mu$ is the mobility),
\be
\frac\gamma{\omega_0}=\frac{0.1}{\mu [10^{6}\ {\rm cm}^{2}/{\rm Vs}]}\sqrt{\frac{\epsilon W[{\rm mm}]}{n_s[10^{11}\ {\rm cm}^{-2}]}}.
\ee
We also introduce the retardation parameter $\alpha$, defined as the ratio of the frequency of the quasi-static 2D plasmon (\ref{2Dp}) with the wavector $q= \pi/W$ to the frequency of light $\pi c/\sqrt{\epsilon}W$ with the same wavector,
\be
\alpha=\frac{\omega_0 \sqrt{\epsilon}W}{\pi c}=\sqrt{\frac{2n_se^2W}{m^\star c^2}};
\label{alpha}
\ee
in the considered case of the uniform dielectric constant the retardation parameter $\alpha$ does not depend on $\epsilon$. Its numerical value is negligibly small for microscopic 2DES (for instance for quantum wires and dots), but can be of order unity for mm-size 2D samples,
\be
\alpha= 0.29\sqrt{n_s[10^{11}\ {\rm cm}^{-2}]\times W[{\rm mm}]}.
\label{alphanum}
\ee
The formula (\ref{alpha}) can be also presented as
\be
\alpha=\frac{\Gamma}{\omega_0},
\ee
where 
\be
\Gamma=\frac{2\pi n_se^2}{m^\star c\sqrt{\epsilon}}
\label{rad}
\ee
is the radiative decay rate in an {\em infinite} sample (see a discussion in \cite{Mikhailov04a}).

We solve the system of equations (\ref{a1})--(\ref{a2}) numerically, restricting ourselves by a finite number $N_h$ of harmonics in the expansions (\ref{jxy}). The matrix equation (\ref{a1})--(\ref{a2}) then has the size $2N_h\times 2N_h$. Convergence of the solution with respect to $N_h$ has been checked. All results shown below have been obtained with $N_h=16$.
For the conductivity tensor $\sigma_{\alpha\beta}$ we have used the Drude model
\be
\sigma_{xx}=\sigma_{yy}=i\frac{n_se^2}{m^\star}\frac{\omega+i\gamma}{(\omega+i\gamma)^2-\omega_c^2},
\ee
\be
\sigma_{xy}=-\sigma_{yx}=\frac{n_se^2}{m^\star}\frac{\omega_c}{(\omega+i\gamma)^2-\omega_c^2}.
\ee

\subsection{Zero magnetic field}

We begin the presentation of results with the case of $B=0$. We assume that the incident electromagnetic wave is linearly polarized, and show results only for the  perpendicular polarization ($E_y^{ext}=0$). The case of the parallel polarization is less interesting for the goals of this work. 

\begin{figure}
\includegraphics[width=8.4cm]{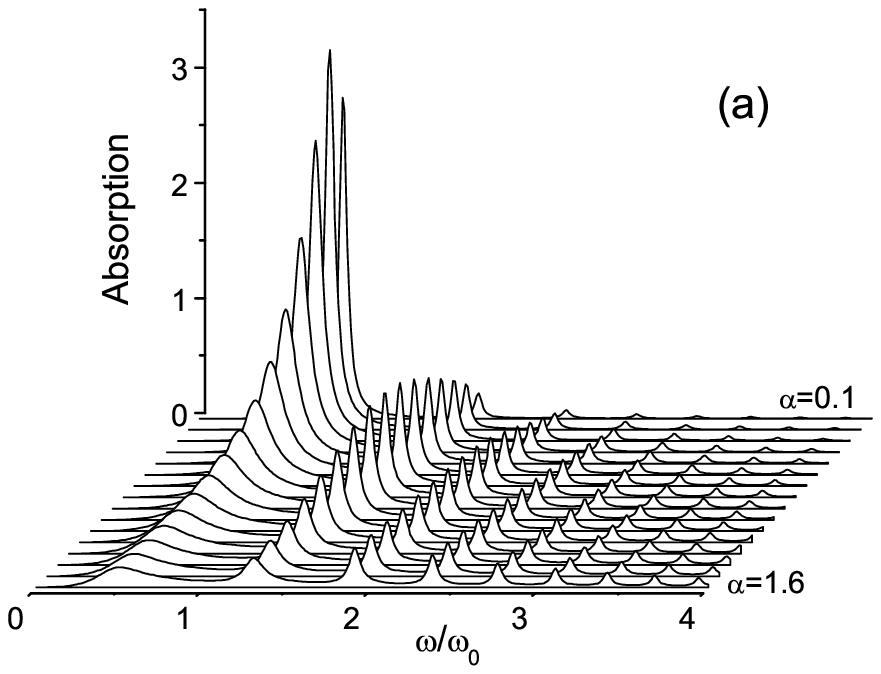}\\
\includegraphics[width=8.4cm]{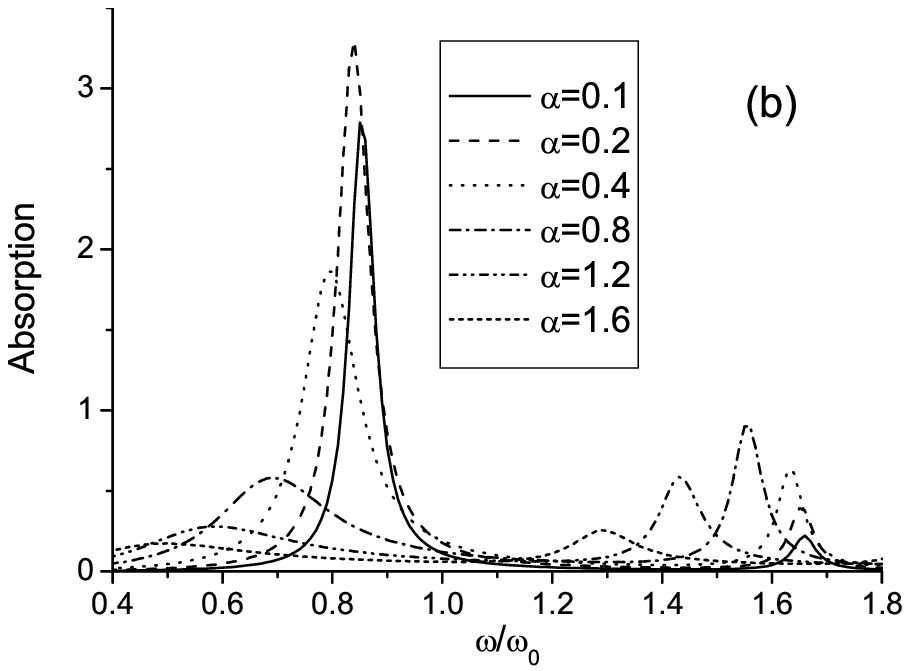}
\caption{Absorption spectra of a 2D stripe at $\gamma/\omega_0=0.05$ and at different retardation parameters. 16 curves are shown in Fig. (a) with $\alpha$ varying from 0.1 up to 1.6 with the step 0.1. Fig. (b) provides a more detailed view on the same spectra around the first and the second modes.}
\label{B0}
\end{figure}

Figure \ref{B0} exhibits the absorption spectra of a stripe as a function of frequency, at $\gamma/\omega_0=0.05$ and a few values of the retardation parameter $\alpha$. One sees that at small $\alpha$  (the quasistatic limit) the absorption spectrum is similar to that of quantum wires: a strong peak corresponding to the fundamental 2D plasmon mode with $q\sim\pi/W$ is accompanied by a number of very weak higher-harmonics peaks, corresponding to $q\sim (2n-1)\pi/W$ with $n=2,3,\dots$. Due to the symmetry of the external electric field $E_x^{ext}(x)=E_x^{ext}=const$, the even 2D plasmon modes with $q\sim 2n\pi/W$ are not excited. When $\alpha$ increases, the fundamental mode first increases in amplitude, reaches its maximum, and then decreases, additionally experiencing a very strong broadening. The behavior of the second mode is similar, but it reaches its maximum at a larger value of $\alpha$. As a result, at a certain $\alpha$ the second-mode peak has a larger amplitude and a smaller linewidth than the first one (see e.g. the curve for $\alpha=0.8$). The third, fourth and all other modes behave similarly, but each subsequent mode reaches its maximum (and becomes the dominant one in the spectrum) at a higher value of the retardation parameter. In general, as seen from the Figure \ref{B0}, at $\alpha\gtrsim 1$ the absorption spectrum has the shape {\em qualitatively different} from that in the quasistatic limit: it consists of many peaks with quite comparable amplitudes, in contrast to the quantum wire/dot spectra, when the higher modes are hardly visible and only the lowest one is dominant. All the modes, as seen from Fig. \ref{B0},  experience quite essential red-shift, in agreement with \cite{Stern67,Kukushkin03a,Kukushkin03b}. 

\subsection{\label{zeroB}Finite magnetic fields}

\begin{figure*}[ht!]
\includegraphics[width=8.4cm]{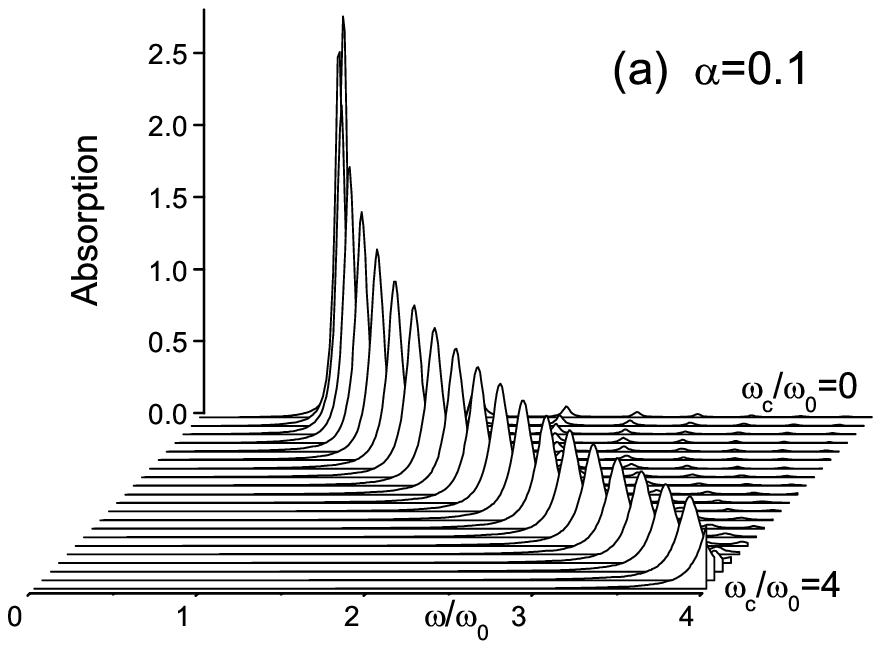}\includegraphics[width=8.4cm]{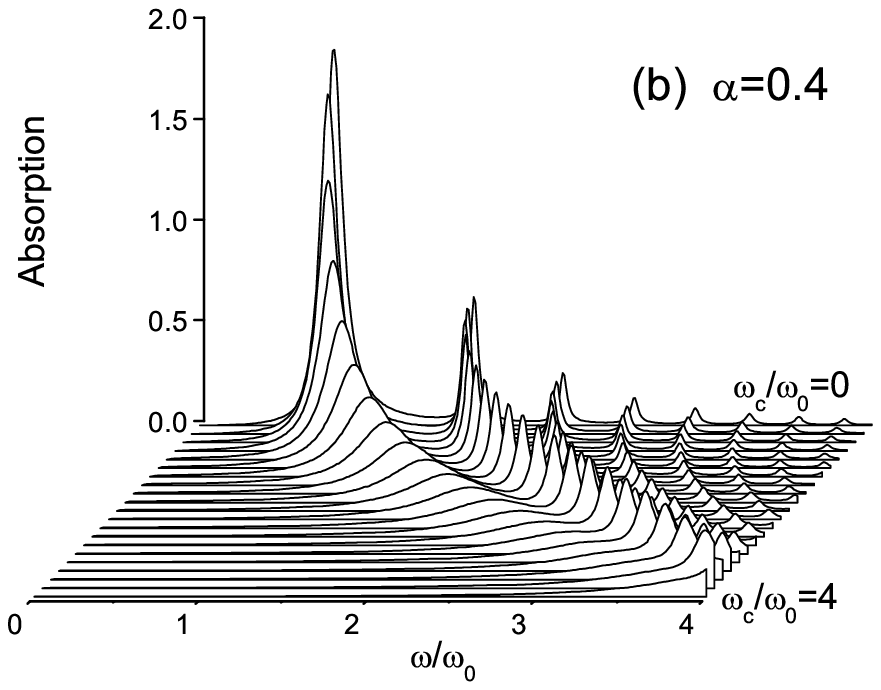}\\
\includegraphics[width=8.4cm]{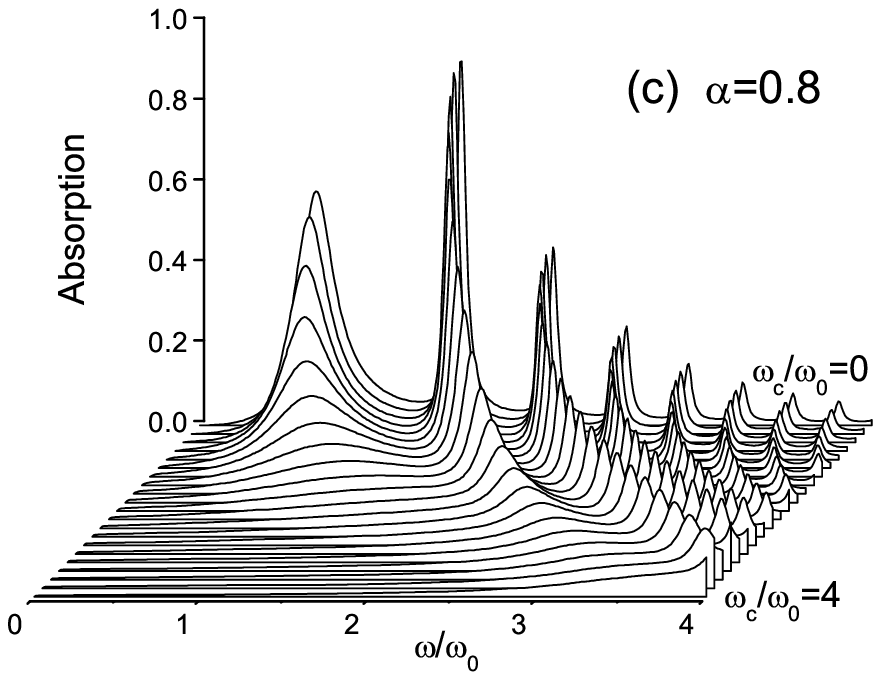}\includegraphics[width=8.4cm]{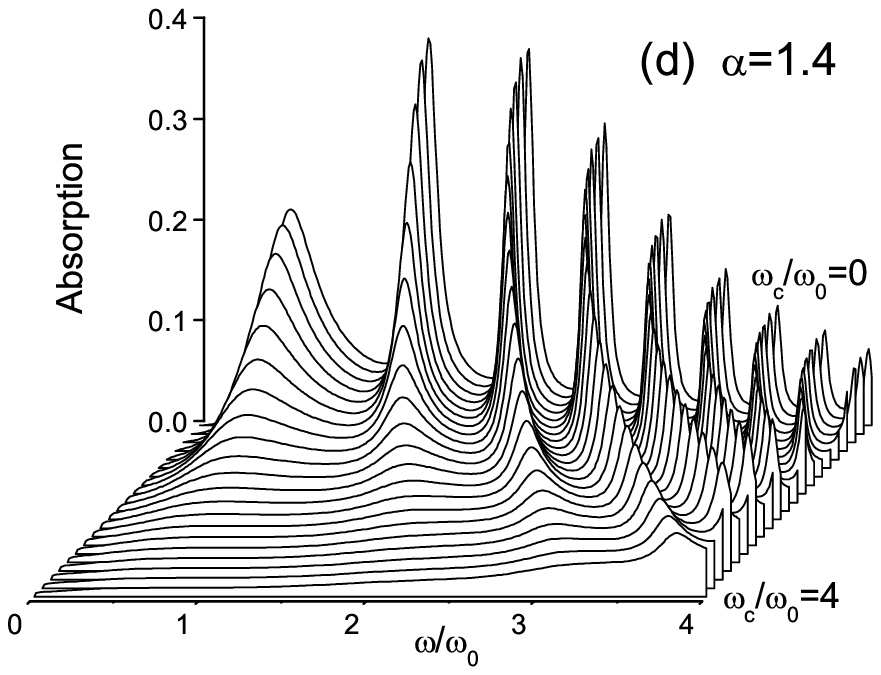}
\caption{Absorption spectra of a 2D stripe as a function of frequency and magnetic field, at $\gamma/\omega_0=0.05$ and at a few representative values of the retardation parameter $\alpha=0.1$, 0.4, 0.8 and 1.4. Each plot contains 21 curves for $\omega_c/\omega_0$ varying from 0 to 4 with the step 0.2.}
\label{B}
\end{figure*}

Figure \ref{B} illustrates the behavior of modes in finite magnetic fields. If $\alpha$ is small, Figure \ref{B}a, the absorption spectrum is, again, very similar to that of quantum wires. The fundamental mode approaches the CR line according to the simple law 
$
\omega_{mp}(B)=\sqrt{\omega^2_p+\omega^2_c}
$,
slightly increasing in the linewidth and decreasing in the amplitude. All the higher modes behave similarly; no one of them becomes stronger than the fundamental mode. Such a behavior was many times observed in experiments on quantum-wire arrays, see e.g.  \cite{Demel88,Demel91}.

If $\alpha$ becomes larger, Fig. \ref{B}b, the lowest, fundamental mode losses its amplitude very quickly approaching the CR line, and the second mode becomes stronger. At even higher frequencies, and at larger values of $\alpha$, Fig. \ref{B}c, the same happens with the second mode: it disappears, approaching and intersecting the CR, and releases its strength to the next mode. Experimentally, this manifests itself as a zigzag behavior of resonances \cite{Kukushkin03a,Kukushkin03b} (we remind that in experiments the absorption was measured as a function of magnetic field at a fixed microwave frequency): when $\omega$ increases, the lower-mode peak gets broader and disappears, intersecting the CR line, while a new, stronger peak, corresponding to the next, higher mode, arises at smaller $B$. At even larger values of $\alpha$, Fig. \ref{B}d, such a relay-race of oscillator strengths between the modes becomes more and more pronounced, especially for higher modes. The frequency of the lowest modes exhibits a very weak magnetic-field dependence at large $\alpha$, Fig. \ref{B}d.

\begin{figure}
\includegraphics[width=8.4cm]{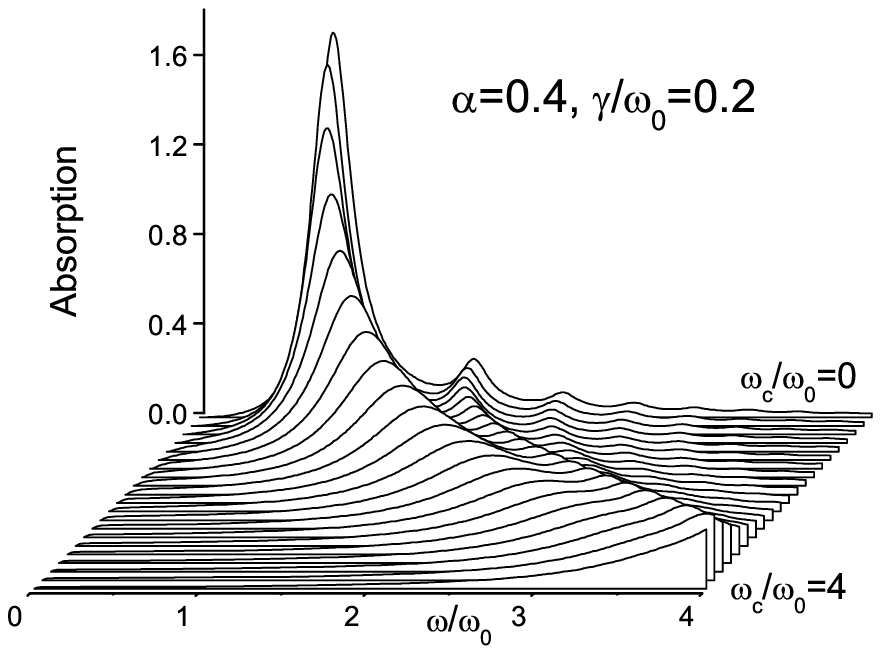}
\caption{Absorption spectra of a 2D stripe as a function of frequency and magnetic field, at $\gamma/\omega_0=0.2$ and $\alpha=0.4$.  }
\label{badsample}
\end{figure}

It should be emphasized that the predicted features can be observed only in samples with a sufficiently high electron mobility. Figure \ref{badsample} exhibits the absorption spectra at $\alpha=0.4$ (like in Figure \ref{B}b), but for a 4 times higher scattering rate ($\gamma/\omega_0=0.2$). One sees that the non-trivial exchange by oscillator strengths between the different modes almost disappears and the whole spectrum reminds now the one shown in Fig. \ref{B}a (with broader linewidths). 

\subsection{Discussion}

As known, in an infinite 2DES the CR has a maximum when the radiative and dissipative decay rates equal to each other. The absorption coefficient in this case (for a wave circularly polarized in the CR direction) has the form \cite{Chiu76,Mikhailov04a}
\be
A_\infty=
\frac{2\gamma\Gamma}{(\omega-\omega_c)^2+(\gamma+\Gamma)^2},
\label{infsample}
\ee
which shows that the CR peak decreases and tends to zero both at $\Gamma/\gamma\to 0$ and $\Gamma/\gamma\to\infty$, and reaches its maximum ($=1/2$) at $\Gamma/\gamma=1$. Similar situation is  the case in 
a finite-width stripe, and it is the competition between the radiative and non-radiative processes that explains the zigzag behavior of modes.

Consider first the case $B=0$, Fig. \ref{B0}. In the quasistatic limit $\alpha\to 0$ the linewidth of the $n$-th mode is determined by $\gamma$, while its amplitudes -- by the radiative decay rate $\Gamma_n$, which gradually decrease with $n$ because of the gradual decrease of the corresponding dipole matrix elements (we stress that $\Gamma_n$ are {\em not identical} to the radiative decay rate $\Gamma$ (\ref{rad}) in an infinite sample). The growth of $\alpha$ leads to the increase of the radiative losses. When the radiative decay rate $\Gamma_n$ of the $n$-th mode becomes equal to the dissipative decay rate $\gamma$, the amplitude of this mode reachs its maximum. As $\Gamma_n$ is a decreasing function of $n$, the absorption maxima for the higher modes are observed at larger values of $\alpha$. This explains the behavior of resonances at $B=0$, Fig. \ref{B0}.

\begin{figure}
\includegraphics[width=8.4cm]{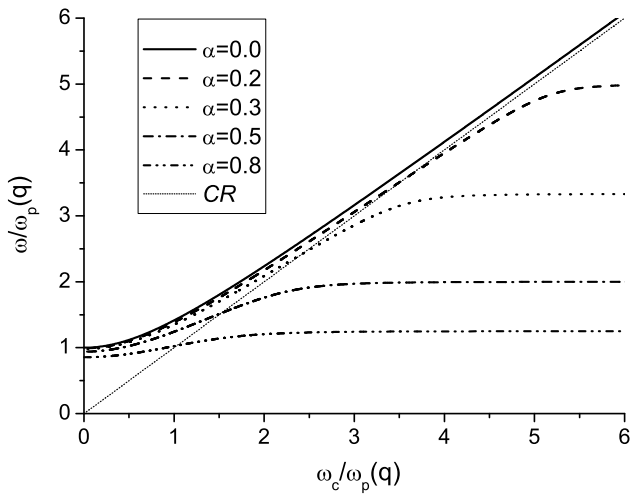}
\caption{The frequency versus magnetic field dependence of magnetoplasmon-polariton modes, as obtained in Ref. \cite{Chiu74}. The units $\omega/\omega_p(q)$, $\omega_c/\omega_p(q)$, and $\alpha^\star=\Gamma/\omega_p(q)$ are used on this plot; for explanations see the text. At large $\omega_c/\omega_p(q)$ the curves tend to the asymptotes $\omega/\omega_p(q)=1/\alpha^\star(q)$, which correspond to the dispersion of light $\omega=cq/\sqrt{\epsilon}$.}
\label{chiu}
\end{figure}

In order to understand the behavior of modes at finite magnetic fields, it is helpful to compare our results with those of Ref. \cite{Chiu74}. Figure \ref{chiu} exhibits the calculated in  \cite{Chiu74} dependence of frequency on magnetic field. As in \cite{Chiu74} the eigen-value problem with a continuous magnetoplasmon wavevector $q$ was considered, we normalize (only on this plot) the frequencies $\omega$ and $\omega_c$ to $\omega_p(q)$, Eq. (\ref{2Dp}), and use the retardation parameter defined as
\be
\alpha^\star\equiv\alpha^\star(q)=\frac{\omega_p(q) \sqrt{\epsilon}}{cq}=\frac{\Gamma}{\omega_p(q)};
\ee
$\omega_p(q)$ and $\alpha^\star(q)$ are reduced to $\omega_0$ and $\alpha$, if $q=\pi/W$. As seen from Fig. \ref{chiu}, the magnetoplasmon dispersion exhibits two different regimes. If $\omega_c/\omega_p\ll 1/\alpha^\star$, it follows the quasistatic behavior $\omega=\sqrt{\omega_p^2+\omega_c^2}$. The interaction of magnetoplasmons with light in this regime is small. If $\omega_c/\omega_p\gtrsim 1/\alpha^\star$, the magnetoplasmon mode intersects the CR line and tends to the asymptote $\omega/\omega_p=1/\alpha^\star$, which corresponds to the dispersion of light $\omega=cq/\sqrt{\epsilon}$. In this regime, the magnetoplasmon strongly interacts with light. In the infinite-sample geometry \cite{Chiu74} the interaction of plasma modes with light thus manifests itself in the change of the dispersion of modes. 

In the finite-size geometry, strong interaction of magnetoplasmons with light in this regime manifests itself in the strong radiative decay of modes. As in a stripe the $n$-th mode corresponds to $q=\pi(2n-1)/W$, the strong interaction regime should be expected at $\omega/\omega_0\simeq (2n-1)/\alpha$. As seen from Fig. \ref{B}, a substantial broadening of the modes is observed, indeed, when the $n$-th mode frequency $\omega/\omega_0$ approaches $(2n-1)/\alpha$. As is also seen from these considerations, the modes with smaller $n$ approach the strong interaction regime and die at lower frequencies, releasing their oscillator strength to the higher modes. This explains the non-trivial zigzag behavior of the magnetoplasmon-polariton resonances, which is seen in Fig. \ref{B} and was experimentally observed in Refs. \cite{Kukushkin03a,Kukushkin03b}. 

We complete our discussion by answering the question, what happens with the spectra of Figure \ref{B} in the limit of a large sample. How the complicated structure of modes, shown in Fig. \ref{B}, is transformed to a simple, single-mode behavior of Eq. (\ref{infsample}) in the limit $W\to\infty$? To answer this question, notice that our frequency unit $\omega_0$ is proportional to $1/\sqrt{W}$ and decreases when $W\to\infty$. This means that the dimensionless dissipation parameter $\gamma/\omega_0$ grows up, and that the multiple-resonance region observed at $\omega\sim \omega_0$ is shifted to smaller frequencies in real (dimensional) units. As seen for example from Fig. \ref{badsample}, in this limit one does get a single-resonance behavior of the absorption spectrum.

\section{conclusions}

As seen from our analysis, the most interesting effects, related to retardation, can be observed in 2DESs under the  condition
\be
\gamma\ll\omega_0\lesssim \Gamma.
\label{cond}
\ee
This implies, first of all, that the parameter $\gamma/\Gamma$ should be small. The condition $\gamma/\Gamma\ll 1$ imposes the upper boundary on the zero-$B$ resistance of 2D samples: it should be much smaller than the wave resistance of vacuum, 377 Ohm (it is so, if the electron mobility in typical GaAs samples exceeds $\sim 10^6$ cm$^2$/Vs). Second, Eq. (\ref{cond}) implies that the typical size of the samples should lie in a certain window, which extends (for typical material parameters) from about 1 mm up to a few cm (dependent on the mobility). Recently published very interesting microwave experiments on the 2DES \cite{Kukushkin03a,Kukushkin03b,Mani02,Zudov03,Dorozhkin03,Yang03,Mani04,Studenikin04,Kukushkin04a,Willett03} have been done in this range of parameters, and hence the phenomena discussed here should be important for their proper interpretation \cite{Mikhailov04a}. 

To summarize, we have theoretically studied microwave absorption spectra of a finite-width 2D stripe, taking into account retardation effects. We have shown that retardation substantially modifies the microwave response of macroscopic 2DESs, especially in the very-high-mobility samples. Our results are in very good agreement with experimental observations, and explain the mysterious zigzag dispersion of magnetoplasmon modes, discovered in \cite{Kukushkin03a,Kukushkin03b}. Further experimental studies of the predicted features in a broader range of parameters would be very interesting.

\begin{acknowledgments}
We are grateful to Igor Kukushkin, Jurgen Smet, Klaus von Klitzing and Detlef Heitmann for  useful discussions of physical issues raised in this paper.
\end{acknowledgments}

\bibliography{../../../../BIB-FILES/mikhailov,../../../../BIB-FILES/lowD,../../../../BIB-FILES/dots,../../../../BIB-FILES/zerores,../../../../BIB-FILES/emp}

\begin{thebibliography}{25}
\expandafter\ifx\csname natexlab\endcsname\relax\def\natexlab#1{#1}\fi
\expandafter\ifx\csname bibnamefont\endcsname\relax
  \def\bibnamefont#1{#1}\fi
\expandafter\ifx\csname bibfnamefont\endcsname\relax
  \def\bibfnamefont#1{#1}\fi
\expandafter\ifx\csname citenamefont\endcsname\relax
  \def\citenamefont#1{#1}\fi
\expandafter\ifx\csname url\endcsname\relax
  \def\url#1{\texttt{#1}}\fi
\expandafter\ifx\csname urlprefix\endcsname\relax\def\urlprefix{URL }\fi
\providecommand{\bibinfo}[2]{#2}
\providecommand{\eprint}[2][]{\url{#2}}

\bibitem[{\citenamefont{Grimes and Adams}(1976)}]{Grimes76}
\bibinfo{author}{\bibfnamefont{C.~C.} \bibnamefont{Grimes}} \bibnamefont{and}
  \bibinfo{author}{\bibfnamefont{G.}~\bibnamefont{Adams}},
  \bibinfo{journal}{Phys. Rev. Lett.} \textbf{\bibinfo{volume}{36}},
  \bibinfo{pages}{145} (\bibinfo{year}{1976}).

\bibitem[{\citenamefont{Allen et~al.}(1977)\citenamefont{Allen, Tsui, and
  Logan}}]{Allen77}
\bibinfo{author}{\bibfnamefont{S.~J.} \bibnamefont{Allen}, \bibfnamefont{Jr.}},
  \bibinfo{author}{\bibfnamefont{D.~C.} \bibnamefont{Tsui}}, \bibnamefont{and}
  \bibinfo{author}{\bibfnamefont{R.~A.} \bibnamefont{Logan}},
  \bibinfo{journal}{Phys. Rev. Lett.} \textbf{\bibinfo{volume}{38}},
  \bibinfo{pages}{980} (\bibinfo{year}{1977}).

\bibitem[{\citenamefont{Theis et~al.}(1978)\citenamefont{Theis, Kotthaus, and
  Stiles}}]{Theis78}
\bibinfo{author}{\bibfnamefont{T.~N.} \bibnamefont{Theis}},
  \bibinfo{author}{\bibfnamefont{J.~P.} \bibnamefont{Kotthaus}},
  \bibnamefont{and} \bibinfo{author}{\bibfnamefont{P.~J.}
  \bibnamefont{Stiles}}, \bibinfo{journal}{Solid State Commun.}
  \textbf{\bibinfo{volume}{26}}, \bibinfo{pages}{603} (\bibinfo{year}{1978}).

\bibitem[{\citenamefont{Theis}(1980)}]{Theis80}
\bibinfo{author}{\bibfnamefont{T.~N.} \bibnamefont{Theis}},
  \bibinfo{journal}{Surf. Sci.} \textbf{\bibinfo{volume}{98}},
  \bibinfo{pages}{515} (\bibinfo{year}{1980}).

\bibitem[{\citenamefont{Heitmann}(1986)}]{Heitmann86}
\bibinfo{author}{\bibfnamefont{D.}~\bibnamefont{Heitmann}},
  \bibinfo{journal}{Surf. Sci.} \textbf{\bibinfo{volume}{170}},
  \bibinfo{pages}{332} (\bibinfo{year}{1986}).

\bibitem[{\citenamefont{Heitmann and Kotthaus}(June 1993)}]{Heitmann93}
\bibinfo{author}{\bibfnamefont{D.}~\bibnamefont{Heitmann}} \bibnamefont{and}
  \bibinfo{author}{\bibfnamefont{J.~P.} \bibnamefont{Kotthaus}},
  \bibinfo{journal}{Physics Today} \textbf{\bibinfo{volume}{46}},
  \bibinfo{pages}{56} (\bibinfo{year}{June 1993}).

\bibitem[{\citenamefont{Allen et~al.}(1983)\citenamefont{Allen, St\"ormer, and
  Hwang}}]{Allen83}
\bibinfo{author}{\bibfnamefont{S.~J.} \bibnamefont{Allen}, \bibfnamefont{Jr.}},
  \bibinfo{author}{\bibfnamefont{H.~L.} \bibnamefont{St\"ormer}},
  \bibnamefont{and} \bibinfo{author}{\bibfnamefont{J.~C.~M.}
  \bibnamefont{Hwang}}, \bibinfo{journal}{Phys. Rev. B}
  \textbf{\bibinfo{volume}{28}}, \bibinfo{pages}{4875} (\bibinfo{year}{1983}).

\bibitem[{\citenamefont{Demel et~al.}(1988)\citenamefont{Demel, Heitmann,
  Grambow, and Ploog}}]{Demel88}
\bibinfo{author}{\bibfnamefont{T.}~\bibnamefont{Demel}},
  \bibinfo{author}{\bibfnamefont{D.}~\bibnamefont{Heitmann}},
  \bibinfo{author}{\bibfnamefont{P.}~\bibnamefont{Grambow}}, \bibnamefont{and}
  \bibinfo{author}{\bibfnamefont{K.}~\bibnamefont{Ploog}},
  \bibinfo{journal}{Phys. Rev. B} \textbf{\bibinfo{volume}{38}},
  \bibinfo{pages}{12732} (\bibinfo{year}{1988}).

\bibitem[{\citenamefont{Demel et~al.}(1990)\citenamefont{Demel, Heitmann,
  Grambow, and Ploog}}]{Demel90}
\bibinfo{author}{\bibfnamefont{T.}~\bibnamefont{Demel}},
  \bibinfo{author}{\bibfnamefont{D.}~\bibnamefont{Heitmann}},
  \bibinfo{author}{\bibfnamefont{P.}~\bibnamefont{Grambow}}, \bibnamefont{and}
  \bibinfo{author}{\bibfnamefont{K.}~\bibnamefont{Ploog}},
  \bibinfo{journal}{Phys. Rev. Lett.} \textbf{\bibinfo{volume}{64}},
  \bibinfo{pages}{788} (\bibinfo{year}{1990}).

\bibitem[{\citenamefont{Demel et~al.}(1991)\citenamefont{Demel, Heitmann,
  Grambow, and Ploog}}]{Demel91}
\bibinfo{author}{\bibfnamefont{T.}~\bibnamefont{Demel}},
  \bibinfo{author}{\bibfnamefont{D.}~\bibnamefont{Heitmann}},
  \bibinfo{author}{\bibfnamefont{P.}~\bibnamefont{Grambow}}, \bibnamefont{and}
  \bibinfo{author}{\bibfnamefont{K.}~\bibnamefont{Ploog}},
  \bibinfo{journal}{Phys. Rev. Lett.} \textbf{\bibinfo{volume}{66}},
  \bibinfo{pages}{2657} (\bibinfo{year}{1991}).

\bibitem[{\citenamefont{Heitmann et~al.}(1992)\citenamefont{Heitmann, Kern,
  Demel, Grambow, Ploog, and Zhang}}]{Heitmann92a}
\bibinfo{author}{\bibfnamefont{D.}~\bibnamefont{Heitmann}},
  \bibinfo{author}{\bibfnamefont{K.}~\bibnamefont{Kern}},
  \bibinfo{author}{\bibfnamefont{T.}~\bibnamefont{Demel}},
  \bibinfo{author}{\bibfnamefont{P.}~\bibnamefont{Grambow}},
  \bibinfo{author}{\bibfnamefont{K.}~\bibnamefont{Ploog}}, \bibnamefont{and}
  \bibinfo{author}{\bibfnamefont{Y.~H.} \bibnamefont{Zhang}},
  \bibinfo{journal}{Surf. Sci.} \textbf{\bibinfo{volume}{267}},
  \bibinfo{pages}{245} (\bibinfo{year}{1992}).

\bibitem[{\citenamefont{Stern}(1967)}]{Stern67}
\bibinfo{author}{\bibfnamefont{F.}~\bibnamefont{Stern}},
  \bibinfo{journal}{Phys. Rev. Lett.} \textbf{\bibinfo{volume}{18}},
  \bibinfo{pages}{546} (\bibinfo{year}{1967}).

\bibitem[{\citenamefont{Chiu and Quinn}(1974)}]{Chiu74}
\bibinfo{author}{\bibfnamefont{K.~W.} \bibnamefont{Chiu}} \bibnamefont{and}
  \bibinfo{author}{\bibfnamefont{J.~J.} \bibnamefont{Quinn}},
  \bibinfo{journal}{Phys. Rev. B} \textbf{\bibinfo{volume}{9}},
  \bibinfo{pages}{4724} (\bibinfo{year}{1974}).

\bibitem[{\citenamefont{Kukushkin
  et~al.}(2003{\natexlab{a}})\citenamefont{Kukushkin, Smet, Mikhailov,
  Kulakovskii, von Klitzing, and Wegscheider}}]{Kukushkin03a}
\bibinfo{author}{\bibfnamefont{I.~V.} \bibnamefont{Kukushkin}},
  \bibinfo{author}{\bibfnamefont{J.~H.} \bibnamefont{Smet}},
  \bibinfo{author}{\bibfnamefont{S.~A.} \bibnamefont{Mikhailov}},
  \bibinfo{author}{\bibfnamefont{D.~V.} \bibnamefont{Kulakovskii}},
  \bibinfo{author}{\bibfnamefont{K.}~\bibnamefont{von Klitzing}},
  \bibnamefont{and}
  \bibinfo{author}{\bibfnamefont{W.}~\bibnamefont{Wegscheider}},
  \bibinfo{journal}{Phys. Rev. Lett.} \textbf{\bibinfo{volume}{90}},
  \bibinfo{pages}{156801} (\bibinfo{year}{2003}{\natexlab{a}}).

\bibitem[{\citenamefont{Kukushkin
  et~al.}(2003{\natexlab{b}})\citenamefont{Kukushkin, Kulakovskii, Mikhailov,
  Smet, and von Klitzing}}]{Kukushkin03b}
\bibinfo{author}{\bibfnamefont{I.~V.} \bibnamefont{Kukushkin}},
  \bibinfo{author}{\bibfnamefont{D.~V.} \bibnamefont{Kulakovskii}},
  \bibinfo{author}{\bibfnamefont{S.~A.} \bibnamefont{Mikhailov}},
  \bibinfo{author}{\bibfnamefont{J.~H.} \bibnamefont{Smet}}, \bibnamefont{and}
  \bibinfo{author}{\bibfnamefont{K.}~\bibnamefont{von Klitzing}},
  \bibinfo{journal}{JETP Letters} \textbf{\bibinfo{volume}{77}},
  \bibinfo{pages}{497} (\bibinfo{year}{2003}{\natexlab{b}}).

\bibitem[{\citenamefont{Mikhailov}(2004)}]{Mikhailov04a}
\bibinfo{author}{\bibfnamefont{S.~A.} \bibnamefont{Mikhailov}},
  \bibinfo{journal}{Phys. Rev. B} \textbf{\bibinfo{volume}{70}},
  \bibinfo{pages}{??} (\bibinfo{year}{2004}),
  \bibinfo{note}{(cond-mat/0405136)}.

\bibitem[{\citenamefont{Mani et~al.}(2002)\citenamefont{Mani, Smet, von
  Klitzing, Narayanamurti, Johnson, and Umansky}}]{Mani02}
\bibinfo{author}{\bibfnamefont{R.~G.} \bibnamefont{Mani}},
  \bibinfo{author}{\bibfnamefont{J.~H.} \bibnamefont{Smet}},
  \bibinfo{author}{\bibfnamefont{K.}~\bibnamefont{von Klitzing}},
  \bibinfo{author}{\bibfnamefont{V.}~\bibnamefont{Narayanamurti}},
  \bibinfo{author}{\bibfnamefont{W.~B.} \bibnamefont{Johnson}},
  \bibnamefont{and} \bibinfo{author}{\bibfnamefont{V.}~\bibnamefont{Umansky}},
  \bibinfo{journal}{Nature} \textbf{\bibinfo{volume}{420}},
  \bibinfo{pages}{646} (\bibinfo{year}{2002}).

\bibitem[{\citenamefont{Zudov et~al.}(2003)\citenamefont{Zudov, Du, Pfeiffer,
  and West}}]{Zudov03}
\bibinfo{author}{\bibfnamefont{M.~A.} \bibnamefont{Zudov}},
  \bibinfo{author}{\bibfnamefont{R.~R.} \bibnamefont{Du}},
  \bibinfo{author}{\bibfnamefont{L.~N.} \bibnamefont{Pfeiffer}},
  \bibnamefont{and} \bibinfo{author}{\bibfnamefont{K.~W.} \bibnamefont{West}},
  \bibinfo{journal}{Phys. Rev. Lett.} \textbf{\bibinfo{volume}{90}},
  \bibinfo{pages}{046807} (\bibinfo{year}{2003}).

\bibitem[{\citenamefont{Dorozhkin}(2003)}]{Dorozhkin03}
\bibinfo{author}{\bibfnamefont{S.~I.} \bibnamefont{Dorozhkin}},
  \bibinfo{journal}{JETP Letters} \textbf{\bibinfo{volume}{77}},
  \bibinfo{pages}{577} (\bibinfo{year}{2003}).

\bibitem[{\citenamefont{Yang et~al.}(2003)\citenamefont{Yang, Zudov, Knuuttila,
  Du, Pfeiffer, and West}}]{Yang03}
\bibinfo{author}{\bibfnamefont{C.~L.} \bibnamefont{Yang}},
  \bibinfo{author}{\bibfnamefont{M.~A.} \bibnamefont{Zudov}},
  \bibinfo{author}{\bibfnamefont{T.~A.} \bibnamefont{Knuuttila}},
  \bibinfo{author}{\bibfnamefont{R.~R.} \bibnamefont{Du}},
  \bibinfo{author}{\bibfnamefont{L.~N.} \bibnamefont{Pfeiffer}},
  \bibnamefont{and} \bibinfo{author}{\bibfnamefont{K.~W.} \bibnamefont{West}},
  \bibinfo{journal}{Phys. Rev. Lett.} \textbf{\bibinfo{volume}{91}},
  \bibinfo{pages}{096803} (\bibinfo{year}{2003}).

\bibitem[{\citenamefont{Mani et~al.}(2004)\citenamefont{Mani, Smet, von
  Klitzing, Narayanamurti, Johnson, and Umansky}}]{Mani04}
\bibinfo{author}{\bibfnamefont{R.~G.} \bibnamefont{Mani}},
  \bibinfo{author}{\bibfnamefont{J.~H.} \bibnamefont{Smet}},
  \bibinfo{author}{\bibfnamefont{K.}~\bibnamefont{von Klitzing}},
  \bibinfo{author}{\bibfnamefont{V.}~\bibnamefont{Narayanamurti}},
  \bibinfo{author}{\bibfnamefont{W.~B.} \bibnamefont{Johnson}},
  \bibnamefont{and} \bibinfo{author}{\bibfnamefont{V.}~\bibnamefont{Umansky}},
  \bibinfo{journal}{Phys. Rev. Lett.} \textbf{\bibinfo{volume}{92}},
  \bibinfo{pages}{146801} (\bibinfo{year}{2004}).

\bibitem[{\citenamefont{Studenikin et~al.}(2004)\citenamefont{Studenikin,
  Potemski, Coleridge, Sachrajda, and Wasilewski}}]{Studenikin04}
\bibinfo{author}{\bibfnamefont{S.~A.} \bibnamefont{Studenikin}},
  \bibinfo{author}{\bibfnamefont{M.}~\bibnamefont{Potemski}},
  \bibinfo{author}{\bibfnamefont{P.~T.} \bibnamefont{Coleridge}},
  \bibinfo{author}{\bibfnamefont{A.~S.} \bibnamefont{Sachrajda}},
  \bibnamefont{and} \bibinfo{author}{\bibfnamefont{Z.~R.}
  \bibnamefont{Wasilewski}}, \bibinfo{journal}{Solid State Commun.}
  \textbf{\bibinfo{volume}{129}}, \bibinfo{pages}{341} (\bibinfo{year}{2004}).

\bibitem[{\citenamefont{Kukushkin et~al.}(2004)\citenamefont{Kukushkin, Akimov,
  Smet, Mikhailov, von Klitzing, Aleiner, and Falko}}]{Kukushkin04a}
\bibinfo{author}{\bibfnamefont{I.~V.} \bibnamefont{Kukushkin}},
  \bibinfo{author}{\bibfnamefont{M.~Y.} \bibnamefont{Akimov}},
  \bibinfo{author}{\bibfnamefont{J.~H.} \bibnamefont{Smet}},
  \bibinfo{author}{\bibfnamefont{S.~A.} \bibnamefont{Mikhailov}},
  \bibinfo{author}{\bibfnamefont{K.}~\bibnamefont{von Klitzing}},
  \bibinfo{author}{\bibfnamefont{I.~L.} \bibnamefont{Aleiner}},
  \bibnamefont{and} \bibinfo{author}{\bibfnamefont{V.~I.} \bibnamefont{Falko}},
  \bibinfo{journal}{Phys. Rev. Lett.} \textbf{\bibinfo{volume}{92}},
  \bibinfo{pages}{236803} (\bibinfo{year}{2004}).

\bibitem[{\citenamefont{Willett et~al.}(2004)\citenamefont{Willett, Pfeiffer,
  and West}}]{Willett03}
\bibinfo{author}{\bibfnamefont{R.~L.} \bibnamefont{Willett}},
  \bibinfo{author}{\bibfnamefont{L.~N.} \bibnamefont{Pfeiffer}},
  \bibnamefont{and} \bibinfo{author}{\bibfnamefont{K.~W.} \bibnamefont{West}},
  \bibinfo{journal}{Phys. Rev. Lett.} \textbf{\bibinfo{volume}{93}},
  \bibinfo{pages}{026804} (\bibinfo{year}{2004}).

\bibitem[{\citenamefont{Chiu et~al.}(1976)\citenamefont{Chiu, Lee, and
  Quinn}}]{Chiu76}
\bibinfo{author}{\bibfnamefont{K.~W.} \bibnamefont{Chiu}},
  \bibinfo{author}{\bibfnamefont{T.~K.} \bibnamefont{Lee}}, \bibnamefont{and}
  \bibinfo{author}{\bibfnamefont{J.~J.} \bibnamefont{Quinn}},
  \bibinfo{journal}{Surf. Sci.} \textbf{\bibinfo{volume}{58}},
  \bibinfo{pages}{182} (\bibinfo{year}{1976}).

\end{thebibliography}

\end{document}